\documentclass[aps,prl,twocolumn,groupedaddress,superscriptaddress,amsfonts,amssymb,amsmath,citeautoscript,a4paper]{revtex4-1}
\pdfoutput=1

\usepackage{amsmath}
\usepackage{amssymb}
\usepackage{graphicx}
\usepackage{subscript}

\usepackage{babel}

\begin{document}

\title{Analytical description of the 1s\textendash exciton linewidth temperature-dependence
in transition metal dichalcogenides}

\author{J. C. G. Henriques}
\affiliation{Department and Centre of Physics, and QuantaLab, University of Minho, Campus of Gualtar, 4710-057, Braga, Portugal}
\affiliation{International Iberian Nanotechnology Laboratory (INL), Av. Mestre Jos{\'e} Veiga, 4715-330, Braga, Portugal}
\affiliation{Center for Nano Optics, University of Southern Denmark, Campusvej 55, DK-5230~Odense~M, Denmark}

\author{N. A. Mortensen}
\affiliation{Center for Nano Optics, University of Southern Denmark, Campusvej 55, DK-5230~Odense~M, Denmark}
\affiliation{Danish Institute for Advanced Study, University of Southern Denmark, Campusvej 55, DK-5230~Odense~M, Denmark}
\affiliation{Center for Nanostructured Graphene, Technical University of Denmark, DK-2800 Kongens Lyngby, Denmark}

\author{N. M. R. Peres}
\affiliation{Department and Centre of Physics, and QuantaLab, University of Minho, Campus of Gualtar, 4710-057, Braga, Portugal}
\affiliation{International Iberian Nanotechnology Laboratory (INL), Av. Mestre Jos{\'e} Veiga, 4715-330, Braga, Portugal}

\begin{abstract}
We obtain an analytical expression for the linewidth of the 1s-exciton as a function of temperature in transition metal dichalcogenides. The total linewidth, as function of temperature, is dominated by three contributions: (i) the radiative decay (essentially temperature independent); (ii) the phonon-induced intravalley scattering; (iii) the phonon-induced intervalley scattering. Our approach uses a variational \emph{Ansatz} to solve the Wannier equation allowing for an analytical treatment of the excitonic problem, including rates of the decay dynamics. Our results are in good agreement with experimental data already present in the literature and can be used to readily predict the value of the total linewidth at any temperature in the broad class of excitonic two-dimensional materials.
\end{abstract}

\maketitle

The optical response of transition metal dichalcogenides (TMDs) is characterized by strong absorption peaks due to excitons formed at the ${\rm K}$ and ${\rm K^{\prime}}$ points of the Brillouin zone~\cite{berghauser2014analytical,wang_colloquium_2018}. The peaks commonly seen in absorption and photoluminescence spectra are associated with the so called bright excitons, in particular to the $s-$states (with vanishing angular momentum) of the $A$ and $B$ series~\cite{berghauser2014analytical,koperski2017optical}. The existence of two distinct series of resonances comes about because of the strong spin-orbit coupling which breaks the spin degeneracy in these systems~\cite{xiao2012coupled}. The effect of spin-orbit coupling is particularly noticeable in the valence band, where in the ${\rm K}$ valley the band with positive spin polarization is shifted downwards relatively to the band with opposite spin. In the conduction band this effect is barely perceptible in MoS\textsubscript{2}, but plays a significant role in WS\textsubscript{2} and WSe\textsubscript{2}~\cite{kormanyos2015k}.

Besides the optically active exciton states, TMDs present a plethora of dark excitons~\cite{malic2018dark,zhang2015experimental,yu2019exploration},
corresponding to exciton states which may be formed, but cannot be directly accessed optically. Examples of these are the $p-$states~\cite{berghauser2018mapping}, the spin-opposed excitons~\cite{dery2015polarization}, and the momentum dark excitons~\cite{selig2016excitonic}. The first ones cannot be directly activated with optical excitation due to angular momentum conservation, however they can be accessed in a pump-probe set-up where the pump laser populates the $1s$ exciton state, and the probe induces transitions between the $1s$ and the $p-$states~\cite{poellmann_resonant_2015}. The spin-opposed excitons are composed of an electron and a hole with opposite spins, and are not optically accessible since the electron-light interaction does not produce the required spin flip. At last, the momentum dark states are composed of an electron and a hole with equal spin, but separated in different momentum points of the Brillouin zone. Due to the small momentum carried by photons, these states are not optically active, however, the momentum mismatch may be overcome and they may be accessed due to additional coupling with phonons~\cite{dey2016optical}.

The rich optical properties of TMDs make them some of the most prominent
materials in the area of nanodevices, with applications ranging from photodetectors, to biosensors and valleytronics~\cite{feierabend2017proposal,ma2021tunable,mueller2018exciton,mak2018light,lv2015transition}. The application of TMDs in devices at finite temperature is limited by the linewidth of the 1s-excitonic line~\cite{cadiz2017excitonic}. Therefore, an understanding of how this linewidth depends both on temperature and on the dielectric media surrounding the two-dimensional (2D) material is key to a complete description of device performance  based on this class of materials. The effect of temperature is mostly dominated by carrier-phonon scattering, and understanding the details of how this scattering mechanism determines the linewidth is essential. Even though this process has already been studied both theoretically and experimentally~\cite{zhang2015experimental,selig2016excitonic}, a simple analytical framework giving insight into the underlying physics is lacking and is necessary.

In this Letter, we derive analytical expressions describing the linewidth of the lowest-lying excitonic resonance in TMDs at finite temperature. We account for the contributions of radiative recombination and exciton-phonon scattering, both intra and intervalley processes. In order to describe these processes we obtain the exciton energies and wave functions from the solution of the Wannier equation using a variational \emph{Ansatz}.
Although the methods and concepts used in this Letter are not entirely new, the combination of all of them together to give an unified analytically picture of the  phonon linewidth in TMDs is the major novelty of this work.

The starting point of our discussion is the introduction of the exciton creation operator. This bosonic operator describes the creation of an exciton with the electron in the point $\xi_{e}$ and the hole in the point $\xi_{h}$ of the Brillouin zone, with center of mass momentum $\mathbf{Q}$ (measured relatively to the $\xi_{e}$ and $\xi_{h}$) and quantum numbers $\nu$ (containing both the principal and angular quantum numbers). This operator is composed of a superposition of electronic operators that annihilate an electron in the valence band and create one in the conduction band, weighted with the Fourier transform of the exciton wave function $\psi_{\nu}(\mathbf{k})$. Then, we assume that the Hamiltonian (containing kinetic and potential energy contributions) is diagonal in this operator, with the eigenvalue $E_{\nu,\mathbf{Q}}=E_{g}+E_{\nu}+\hbar^{2}\mathbf{Q}^{2}/2M$, where $E_{g}$ is the band gap, $E_{\nu}$ the exciton binding energy, and $\hbar^{2}\mathbf{Q}^{2}/2M$ the kinetic energy of the center of mass. Afterwards we compute the commutator of the Hamiltonian with the exciton operator using its representation in terms of exciton operators and in terms of electron operators. Demanding the equality of both results, the Bethe--Salpeter equation (BSE) follows, whose solution yields the binding energies and wave functions of the excitonic problem. The interaction potential in the electronic Hamiltonian is treated within the Rytova--Keldysh formalism~\cite{keldysh1979coulomb,rytova1967,cudazzo2011dielectric}, which crucially accounts for the non\textendash local screening in the TMD monolayer. Fourier transforming the BSE (see SM) we find the Wannier equation~\cite{haug2009quantum}, a differential equation in real space reading
\begin{equation}
-\frac{\hbar^{2}}{2\mu_{\xi_{e}\xi_{h}}}\nabla^{2}\psi_{\nu}(\mathbf{r})+V_{{\rm RK}}(\mathbf{r})\psi_{\nu}(\mathbf{r})=E_{\nu}\psi_{\nu}(\mathbf{r}),\label{eq:Wannier eq}
\end{equation}
where $\mu_{\xi_{e}\xi_{h}}$ is the reduced mass of the electron-hole
pair, with the former and the latter located in the $\xi_{e}$ and
$\xi_{h}$ valley, respectively. In Eq.~(\ref{eq:Wannier eq}), $V_{{\rm RK}}(\mathbf{r})$ is the Rytova--Keldysh potential and $\psi_{\nu}(\mathbf{r})$ the exciton wave function in real space. The Rytova--Keldysh potential follows from the solution of the Poisson equation for a point charge in a thin dielectric and reads~\cite{keldysh1979coulomb,rytova1967}
\begin{equation}
V_{{\rm RK}}(\mathbf{r})=-\frac{\pi}{2r_{0}}\left[H_{0}\left(\frac{\kappa r}{r_{0}}\right)-Y_{0}\left(\frac{\kappa r}{r_{0}}\right)\right],
\end{equation}
where $\kappa$ is the mean dielectric constant of the media above and bellow the TMD layer, and $r_{0}$ is a material parameter which can be macroscopically associated with a screening length, and is microscopically related with the polarizability of the monolayer; $H_{0}$ is the Struve function and $Y_{0}$ the Bessel function of the second kind, both of order zero. The difference between momentum bright and momentum dark excitons lies solely on the value of the reduced mass, which should be computed with the adequate effective masses. To obtain the location of the exciton states the structure of the electronic bands must also be know. In our study we used the band parameters found from \emph{ab initio} calculations in Ref.~\onlinecite{kormanyos2015k}. When we compare our results (see SM) with others found in the literature, either by numerically solving the BSE~\cite{malic2018dark}, or using density-functional theory (DFT) calculations~\cite{deilmann2019finite}, a good agreement is found. We stress that the goal of our approach is not so much the quantitative accuracy, but rather the possibility of obtaining qualitative and physically transparent results using simpler techniques which can be explored analytically.

While the direct solution of the BSE requires a delicate numerical diagonalization, Eq.~(\ref{eq:Wannier eq}) can be solved with a variational approach~\cite{Mauricio2020}, which allows us to develop an analytical framework gaining insight into the underlying physics. Different variational \emph{Ans{\"a}tze} can be employed to solve this problem; here we opt to use the simplest one, a single evanescent exponential $\psi_{1s}(\mathbf{r})=\sqrt{2/(\pi a^{2})}e^{-r/a}$,
similar to the wave function of the Hydrogen atom. The value of $a$ is determined from the minimization of the energy in Eq.~(\ref{eq:Wannier eq}). The comparison of this variational approach with the exact diagonalization of the Hamiltonian~\cite{henriques2019optical} is given in the SM. Following the scaling procedure to a dimensionless representation proposed in Ref.~\onlinecite{pedersen_exciton_2016}, we can conveniently describe the excitonic problem with a single effective parameter $\tilde{r}_{0}=r_{0}\mu/\kappa^{2}$ instead of the three independent ones we currently have. Noting that in the usual TMDs one finds $r_{0}\mu=30-40$ atomic units~\cite{pedersen_exciton_2016,olsen2016simple}, and performing a power-law fit (including a deviation term) of $a$ vs $\tilde{r}_{0}$ in the region of interest of the parameter space (with $\kappa\in[1,5]$), we realize that $a$ is accurately described by (see SM)
\begin{equation}
\frac{a}{a_{B}}\backsimeq\frac{\kappa}{\mu/m_{0}}a_{0}+\sqrt{\frac{r_{0}/a_{B}}{\mu/m_{0}}},\label{eq:a fit}
\end{equation}
where $a_{0}=0.4$ is determined by the fitting process, $m_{0}$ is the bare electron mass, and $a_{B}$ is the Bohr radius. If the fit had been performed in a region corresponding to smaller values of $r_{0}$, and consequently smaller $\tilde{r}_{0}$, the value of $a_{0}$ would become ever closer to $0.5$, just like in the 2D Hydrogen atom, and in agreement with the limiting case $\lim_{r_{0}\rightarrow0}V_{{\rm RK}}(r)=V_{{\rm Coulomb}}(r)$. The fact that in the relevant region of parameters for TMDs the value of $a_{0}$ differs from 0.5 reflects the non-Hydrogenic nature of excitons in these 2D systems; the term $\sqrt{r_{0}/\mu}$ further enhances this difference.

Now that the equation governing the excitonic problem was found, and a simple variational approach of solving it was introduced, we can move on to the computation of the linewidth of the $1s-$exciton in TMDs. Starting with the radiative linewidth~\cite{palummo2015exciton}, and using Fermi's golden rule, we compute the optical matrix element $\langle GS;1_{\mathbf{q}}|\mathbf{E}\cdot\mathbf{r}|1s,\mathbf{Q}=0\rangle$ in the dipole approximation, where the final state corresponds to the excitonic vacuum, $|GS\rangle$, with an extra photon with momentum $\mathbf{q}$ in the light field. Quantizing $\mathbf{E}\cdot\mathbf{r}$ in terms of photon and exciton operators, and carrying out the calculation (see SM) one finds
\begin{equation}
\gamma_{{\rm rad}}=\frac{8\pi}{\kappa}\alpha\left(\hbar v_{F}\right)^{2}\frac{E_{g}+E_{\nu}}{\left(E_{g}^{\tau,s_{z}}\right)^{2}}\psi_{\nu}^{2}(\mathbf{r}=0),\label{eq:gamma rad}
\end{equation}
where $\alpha\sim1/137$ is the fine-structure constant, $v_{F}$
is the Fermi velocity (which may be obtained from first-principles calculations), $\tau=\pm1$ and $s_{z}=\pm1$ are the valley (K/K') and spin (up/down) indices respectively, and $E_{g}^{\tau,s_{z}}$ is the noninteracting band gap, which is spin and valley dependent and is obtained from first-principles calculations~\cite{kormanyos2015k}; the dependence of $\psi_{\nu}^{2}(\mathbf{r}=0)$ on $\kappa$, $\mu$ and $r_{0}$ follows from Eq.~(\ref{eq:a fit}). From Eq.~(\ref{eq:gamma rad}) we expect the radiative linewidth to increase as the material's band gap widens, and to decrease approximately according to a power law as the surrounding dielectric screening increases. Although our expression is independent of the temperature, the radiative linewidth may vary with it as a result of modifications in the band structure. This, however, is beyond the scope of the present work.For the most common TMDs we find $\gamma_{{\rm rad}}\sim5$\,meV, in agreement with other independent results~\cite{palummo2015exciton,scuri2018large}.

In order to describe the linewidth originating from exciton-phonon coupling one must consider the mechanisms responsible for phonon-driven carrier scattering. In the present discussion we will restrict ourselves to the deformation potential framework~\cite{kaasbjerg2012phonon,kaasbjerg2013acoustic,li2013intrinsic,jin2014intrinsic}, which is expected to give the main contribution for exciton-phonon scattering in TMDs. For intravalley acoustic scattering we consider a first-order deformation potential and a linear energy dispersion, that is, we take a Debye approach. For intravalley optical scattering and all intervalley processes we consider a zero-order deformation potential and constant dispersion relations, that is, an Einstein approach. The values of the deformation potentials and the phonon energies are obtained from DFT calculations, as the ones of Ref.~\onlinecite{jin2014intrinsic}. The exciton-phonon coupling is defined as the individual carrier-phonon coupling times the overlap of the wave functions of the initial and final exciton states in momentum space (see SM). %a sum in momentum space of a product of the wave functions of the initial and final exciton states (see SM).
Since in intravalley processes the transferred momentum is small and negligible, this sum effectively corresponds to the overlap of wave functions near the same point in the Brillouin zone, and may be approximated to 1; thus, the exciton-phonon coupling element directly follows from the individual carrier-phonon couplings (see SM).

\begin{figure}
\centering{}\includegraphics{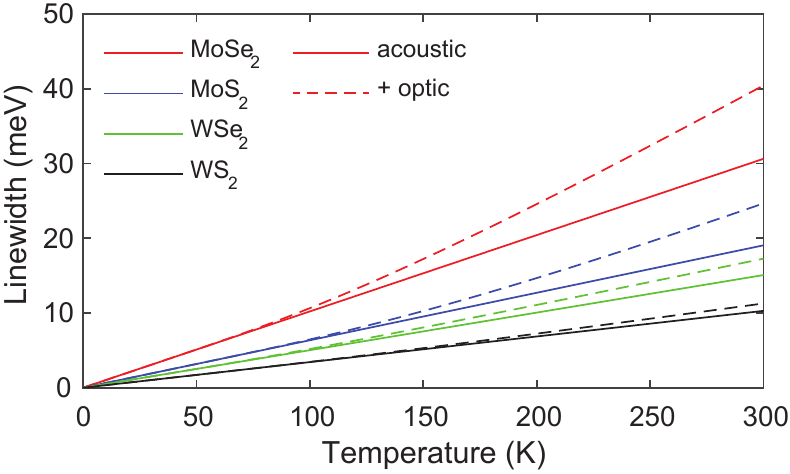}\caption{\label{fig:Intravalley linewidth}Intravalley scattering induced linewidth for four different TMDs. The solid lines represent the contribution from acoustic phonons, and the dashed lines represent the combined contribution of acoustic and optical phonons. The parameters of Ref.~\onlinecite{jin2014intrinsic} were used.}
\end{figure}

To compute the intravalley scattering contribution to the linewidth we once again turn to Fermi's golden rule, and for simplicity assume that intravalley scattering conserves the quantum numbers of the exciton (a reasonable assumption considering the large energy required in, e.g. the $1s\rightarrow2s$ transition). Focusing our study on the optically bright states, $\mathbf{Q}=0$, only processes involving scattering through absorption of phonons are possible. Applying Fermi's golden rule we find
\begin{subequations}
\begin{equation}
\gamma_{{\rm intra,\:ac}}\approx\frac{\left|\Xi^{(e)}-\Xi^{(h)}\right|^{2}}{\rho v_{{\rm ac}}^{2}}\frac{M}{\hbar^{2}}k_{B}T,
\end{equation}
and
%
%\begin{equation}
%\gamma_{{\rm intra,\:op}}=\frac{\left|D^{0,(e)}-D^{0,(h)}\right|^{2}}{2\rho\hbar\omega_{{\rm op}}}\frac{M}{\exp\left(\frac{\hbar\omega_{{\rm op}}}{k_{B}T}\right)-1},
%\end{equation}
\begin{equation}
\gamma_{{\rm intra,\:op}}= M \frac{\left|D^{0,(e)}-D^{0,(h)}\right|^{2}}{2\rho\hbar\omega_{{\rm op}}} n(\hbar\omega_{\rm{op}}),
\end{equation}
\end{subequations}
for the intravalley process due both to acoustic (ac) and optical (op) phonons. Here $\Xi^{(e/h)}$ and $D^{0,(e/h)}$ are the first-order acoustic deformation potential and the zero-order optical deformation potential for electron/hole-phonon intravalley scattering, $n(\hbar\omega)$ is the Bose-Einstein distribution, $v_{{\rm ac}}$ is the speed of sound in the TMD, $\hbar\omega_{{\rm op}}$ is the optical phonon energy, $M=m_{e}+m_{h}$, $\rho$ is the mass density of the monolayer~\cite{shree2018observation}, and $k_{B}$ is the Boltzmann's constant. As expected~\cite{li2013intrinsic,jin2014intrinsic} we observe that the intravalley linewidth from acoustic phonons increases linearly with temperature. The contribution from optical phonons presents the same temperature dependence as the Bose--Einstein distribution function. From DFT calculations one finds that the values of the deformation potential make scattering with acoustic phonons the dominant mechanism of the two~\cite{jin2014intrinsic}, and thus it is expected that the total intravalley scattering linewidth increases linearly with temperature with a small exponential correction. Also, we note that both contributions are insensitive to dielectric screening from the environment. In Fig.~\ref{fig:Intravalley linewidth} we depict the intravalley linewidth as a function of temperature for MoS\textsubscript{2}, MoSe\textsubscript{2}, WS\textsubscript{2} and WSe\textsubscript{2}. We observe that this linewidth takes higher values for molybdenum (Mo) based TMDs than for tungsten (W) based ones, because of the higher effective masses and deformation potentials of the former. Moreover, TMDs with the selenium (Se) are associated with larger broadenings than the ones with sulfur (S); this is due to the higher values of $\rho v_{{\rm {\rm ac}}}$ and $\hbar\omega_{{\rm op}}$, suppressing $\gamma_{{\rm intra,\:ac}}$ and $\gamma_{{\rm intra,\:op}}$, respectively, in TMDs with sulfur.

Now, let us consider the contribution from intervalley scattering to the temperature dependence of the linewidth. For simplicity we will consider that only the electron is scattered, going from the ${\rm K}$ valley to other points of the Brillouin zone, while the hole remains in the ${\rm K}$ valley. This approximation is not expected to have a significant impact on the final result, since the latter is a less efficient scattering process due to the large momentum involved and the smaller values of the deformation potentials~\cite{jin2014intrinsic}. The efficiency of intervalley scattering is directly related to the electronic band structure, and the relative position of the exciton energy levels in different points of the Brillouin zone. In molybdenum based TMDs, where the lowest lying exciton state is located at the ${\rm K}$ valley, we do not expect intervalley processes to play a significant role, since the energetically more favorable scattering event ${\rm K}\rightarrow{\rm K}'$ is rather inefficient. However, the situation is significantly different in tungsten based TMDs, where the conduction band presents a satellite minimum halfway along the path from the vertices of the Brillouin zone to its center, the $\Lambda$ valley. In this region the effective masses are larger than in the ${\rm K}$ valley, leading to more tightly bound excitons. As a consequence, in these TMDs a momentum dark state at the $\Lambda$ valley appears bellow the optically active exciton. Since scattering events of the type ${\rm K}\rightarrow\Lambda$ are energetically favorable, and the required momentum transfer is half of that required in ${\rm K}\rightarrow{\rm K}'$ processes, it is expected that intervalley scattering offers a significant contribution to the total linewidth in tungsten based TMDs.

To compute the intervalley linewidth we return to Fermi's golden rule, modifying the interaction Hamiltonian to explicitly include the valley information of the carriers. The explicit form of the exciton-phonon coupling for intervalley scattering is given in the SM. Just as in the case of intravalley scattering events, the exciton-phonon coupling is given by the product of the carrier-phonon coupling with a sum in momentum space of the wave functions of the initial and final exciton states, which now refer to separated points in the Brillouin zone due to the nature of intervalley scattering. This time around, due to the large transferred momentum this sum cannot be simply approximated to 1, and needs to be explicitly computed. Since we have mapped the problem of both bright and dark excitons to the Wannier equation, which can be solved with a simple variational \emph{Ansatz}, the sum in momentum space can still be computed analytically (see SM). Thus, the nonradiative decay rate associated with the intervalley scattering of an electron from the ${\rm K}$ to the $\xi_{f}$ valley, with the hole remaining in the ${\rm K}$ valley (scattering of a ${\rm KK}$ exciton to a K$\xi_{f}$ one), reads:
\begin{align}
 & \gamma_{{\rm inter,}{\rm K}\xi_{f}}\approx\sum_{\lambda,\pm}w\frac{\left|D_{{\rm K}\rightarrow\xi_{f}}^{0,\lambda,(e)}\right|^{2}}{\rho\hbar\omega_{\mathbf{j},\lambda}}\frac{8M_{\xi_{f}}a_{{\rm K}}^{2}a_{\xi_{f}}^{2}\left(a_{{\rm K}}+a_{\xi_{f}}\right)^{2}}{\left[\left(a_{{\rm K}}+a_{\xi_{f}}\right)^{2}+\beta_{\xi_{f}}^{2}a_{{\rm K}}^{2}a_{\xi_{f}}^{2}\mathbf{j}^{2}\right]^{3}}\nonumber \\
 & \times\left[\frac{1}{2}\pm\frac{1}{2}+n(\hbar\omega_{\mathbf{j},\lambda})\right]\Theta\left(-\Delta_{{\rm K}\xi_{f}}\mp\hbar\omega_{\mathbf{j},\lambda}\right),\label{eq:gamma inter}
\end{align}
where the sums are over the phonon modes $\lambda$ (acoustic and optical), and the emission/absorption $(+/-)$ of phonons and $w$ is the degeneracy factor of the $\rm K\rightarrow\xi_{f}$ scattering event; $\mathbf{j}$ is the momentum given by ${\rm K}-\xi_{f}$, $M_{\xi_{f}}=m_{e,\xi_{f}}+m_{h,{\rm K}}$, is the translational mass of the ${\rm K}\xi_{f}$ exciton, and $\beta_{\xi_{f}}=m_{h,{\rm K}}/M_{\xi_{f}}$. The Heaviside function, $\Theta(x)$, sets the threshold for energetically
allowed scattering processes, with $\Delta_{{\rm K}\xi_{f}}=\Delta_{{\rm K}\xi_{f}}^{{\rm CB}}+\Delta_{{\rm K}\xi_{f}}^{{\rm binding}}$ the energy difference of the conduction band edge at the ${\rm K}$ and $\xi_{f}$ valleys ($\Delta_{{\rm K}\xi_{f}}^{{\rm CB}}$) plus the difference of the binding energy of the  ${\rm KK}$ and K$\xi_{f}$ excitons
($\Delta_{{\rm K}\xi_{f}}^{{\rm binding}}$); $a_{\xi_{f}}(a_{{\rm K}})$ is the variational parameter associated with the wave function of the ${\rm K}\xi_{f}$ (KK) exciton. The dependence of $a$ with the physical parameters $\mu_{\xi_{e},\xi_{h}}$, $\kappa$ and $r_{0}$ is given by Eq.~(\ref{eq:a fit}).

\begin{figure}[h]
\centering{}\includegraphics{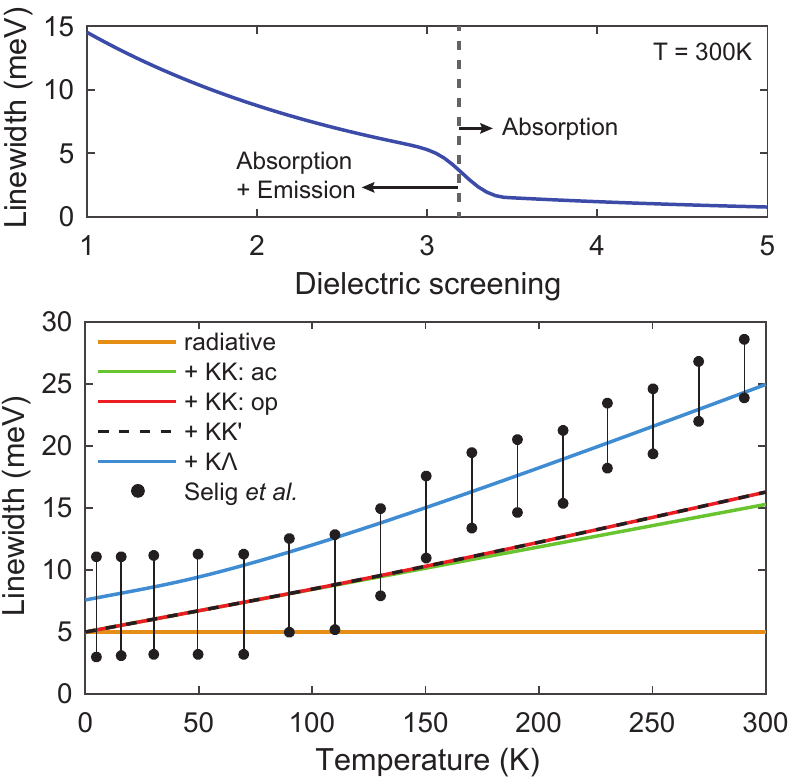}\caption{\label{fig:WS2 total}(Top) Nonradiative linewidth as a function of the dielectric screening at $T=300$ K. (Bottom) Different contributions to the total linewidth of WS\protect\textsubscript{2} as a function of temperature. Contributions labeled by $\rm KK$ refer to intravalley scattering while the others refer to intervalley contributions. The experimental points were taken from Ref.~\onlinecite{selig2016excitonic}, while parameters of Refs.~\onlinecite{kormanyos2015k,jin2014intrinsic} were used.}
\end{figure}

The appearance of the transferred momentum $\mathbf{j}$ in the denominator of Eq.~(\ref{eq:gamma inter})contributes to the suppression of intervalley processes such as ${\rm K}\rightarrow{\rm K}'$ when compared to ${\rm K}\rightarrow\Lambda$.
This is further enhanced by the deformation potential $D_{{\rm K}\rightarrow\xi_{f}}^{0,\lambda,(e)}$, which is approximately 7 times larger in ${\rm K}\rightarrow\Lambda$ processes than in ${\rm K\rightarrow K'}$. Considering that in Eq.~(\ref{eq:gamma inter}) $\gamma_{{\rm inter}}$ is proportional to the square of the deformation potential, it is clear that the contribution from ${\rm K}\rightarrow\Lambda$ is by far the dominant one. Moreover, transitions involving acoustic phonons have larger deformation potentials and thus are more relevant than the ones assisted by optical phonons.
If $\Delta_{{\rm K}\xi_{f}}<0$, which is more likely to occur in tungsten based TMDs due to the structure of its conduction band, then intervalley scattering is highly favorable and can occur via absorption or emission of phonons. On the other hand, if $\Delta_{{\rm K}\xi_{f}}>0$ then only absorption scattering channels are available, and even those
may be suppressed depending on how $\Delta_{{\rm K}\xi_{f}}$ compares with the phonon energies. Contrary to intravalley scattering, the contribution from intervalley processes depends on the dielectric screening from the environment, owning to its dependence on the variational parameter $a$. Analyzing Eq.~(\ref{eq:gamma inter}) combined with Eq.~(\ref{eq:a fit}) we predict a decrease of linewidth with increasing dielectric screening roughly following a power law of sixth order. Furthermore, as the dielectric screening increases, the binding energies decrease, and the ${\rm K}\Lambda$ excitons gets closer to the optically bright one, leading to less energetically favorable intervalley processes,
which results in the suppression of scattering channels and consequent decrease of the linewidth. This effect is depicted in Fig.~\ref{fig:WS2 total}, where we observe a sudden decrease in the nonradiative linewidth due to the suppression of the emission scattering channels. For higher values of the substrate dielectric constant even the absorption scattering
channels could be suppressed, however for such values of screening the intervalley linewidth becomes almost insignificant even before the suppression of the channel. Band gap engineering, and changes in the effective masses may also be used to enhance (or suppress) different scattering channels.

Now that all the necessary analytical expressions were derived, we study the specific case of the variation of the $1s-$exciton linewidth in WS\textsubscript{2} with temperature, and compare our theoretical prediction with the experimental data of Ref.~\onlinecite{selig2016excitonic}; in Fig.~\ref{fig:WS2 total} a good agreement between theory and experiment is seen. We note that the total linewidth is composed of a constant radiative term of approximately 5 meV, combined with the almost linear contribution from intravalley scattering (cf. Fig.~\ref{fig:Intravalley linewidth}),
and an exponentially increasing intervalley term. These contributions amount to a total linewidth at room temperature of approximately 25\,meV. Moreover, the intervalley contributions stems almost entirely from ${\rm K}\rightarrow\Lambda$ events, as a consequence of the low efficiency of the ${\rm K}\rightarrow{\rm K}'$ scattering events (something similar would happen with the ${\rm K\rightarrow\Gamma}$ hole scattering). In the low temperature region we observe that, contrary
to the intravalley linewidth, the intervalley term is finite due to the efficient process of scattering with phonon emission. As the temperature increases so do the intra and intervalley contributions, presenting similar magnitudes at room temperature. A similar calculation for WSe\textsubscript{2} yields an identical result, with a slightly larger linewidth, as a consequence of the more favorable intravalley scattering processes (cf. Fig.~\ref{fig:Intravalley linewidth}). For MoS\textsubscript{2} and MoSe\textsubscript{2}, where intervalley processes are less efficient the total linewidth is basically given by the radiative and intravalley contributions.

In summary, solving the Wannier equation with a variational \emph{Ansatz} we obtained the binding energies and wave functions of bright and dark excitons. Then, employing Fermi's golden, rule we derived analytical expressions describing the radiative, intravalley and intervalley contributions to the total linewidth of the lowest-lying exciton resonance
in different TMDs at finite temperature. Our theoretical prediction is in good agreement with experimental data. The derived expressions, combined with parameters computed from DFT calculations, allow %qualitative insight and 
for easily accessible estimates of the different contributions for the linewidth at any temperature value, and give insight on how these quantities depend on the material and environment parameters.

N.~M.~R.~P acknowledges support by the Portuguese Foundation for Science and Technology (FCT) in the framework of the Strategic Funding UIDB/04650/2020. J.~C.~G.~H. acknowledges the Center of Physics for a grant funded by the UIDB/04650/2020 strategic project and POCI-01-0145-FEDER-028887. N.~M.~R.~P. acknowledges support from the European Commission through the project ``Graphene-Driven Revolutions in ICT and Beyond'' (Ref. No. 881603, CORE 3), COMPETE 2020, PORTUGAL 2020, FEDER and the FCT through projects POCI-01-0145-FEDER-028114, PTDC/NAN-OPT/29265/2017. N.~A.~M. is a VILLUM Investigator supported by VILLUM FONDEN (Grant No.~16498). The Center for Nanostructured Graphene is sponsored by the Danish National Research Foundation (Project No.~DNRF103).

\bibliographystyle{aapmrev4-2}
%aapmrev4-2.bst 2019-01-14 (MD) hand-edited version of aapmrev4-1.bst
%Control: key (0)
%Control: author (8) initials jnrlst
%Control: editor formatted (1) identically to author
%Control: production of article title (-1) disabled
%Control: page (0) single
%Control: year (1) truncated
%Control: production of eprint (0) enabled
%

\end{document}